\documentclass[a4paper,12pt]{article}
\usepackage{amssymb}

\begin{document}

\title{Scalar Field as Dark Matter and Machian Cosmological Solution in the
Generalized Scalar-Tensor Theory of Gravitation}
\author{A. Miyazaki \thanks{
Email: miyazaki@loyno.edu, miyazaki@nagasakipu.ac.jp} \vspace{3mm} \\
\textit{Department of Physics, Loyola University, New Orleans, LA 70118} \\
and \\
\textit{Faculty of Economics, Nagasaki Prefectural University} \\
\textit{Sasebo, Nagasaki 858-8580, Japan}}
\date{\vfill}
\maketitle

\begin{abstract}
The Machian cosmological solution satisfying $\phi =O(\rho /\omega )$ in the
generalized scalar-tensor theory of gravitation with the varying
cosmological constant is summarized. The scalar field $\varphi $ with the
exponential potential is introduced as dark matter and the barotropic
evolution of matter in the universe is discussed. As the universe expands,
the coefficient $\gamma $ of the equation of state approaches to $-1/3$ and
the coupling function $\omega (\phi )$ diverges to $-\infty $. \newline
\newline
\textbf{PACS numbers: 04.50.+h, 98.80.-k }
\end{abstract}

\newpage

It is believed that the field equations of the Brans-Dicke theory \cite{1)}
coincide with those of the Einstein theory with the same energy-momentum
tensor when the coupling parameter $\omega $ goes to the infinity, but this
is generally not true \cite{2)}-\cite{4)}. We proposed the postulate in the
Machian point of view: The scalar field of a proper cosmological solution
should have the asymptotic form $\phi =O(\rho /\omega )$ when $\omega $ is
large enough and should converge to zero in the continuous limit $\rho
/\omega \rightarrow 0$ (the Machian solution) \cite{4)}. We proved that the
Machian solution necessarily satisfies the relation $GM/R=const$ for the
homogeneous and isotropic universe \cite{5)}.

We extended the Machian solution to the case of a perfect fluid with the
negative pressure \cite{6)} (see also \cite{15)}). This solution shows that
the time-variation of the gravitational constant vanishes in the limit of $%
\gamma =-1/3$ and suggests that the coupling parameter $\omega $ of the
Brans-Dicke scalar field should vary in time. We investigated various
possibilities of the coupling function $\omega (\phi )$ in the generalized
scalar-tensor theory of gravitation \cite{7)}-\cite{9)} and obtained $\omega
(\phi )=\eta /(\xi -2)$ to support the reasonable Machian solution \cite{10)}%
. In this cosmological model, the coefficient of the expansion parameter and
the scalar field $\phi $ converge to constants respectively as the parameter 
$\xi \rightarrow 2$ ($\gamma \rightarrow -1/3$). When $\xi \rightarrow 2$,
the coupling function $\omega (\phi )$ diverges to the minus infinity and
the gravitational constant approaches dynamically to the constant $G_{\infty
}$. Moreover, if we require the particular form of the cosmological
constant, the similar Machian cosmological solution exists in the
generalized scalar-tensor theory of gravitation and the universe shows the
slowly accelerating expansion at present \cite{11)}, which is compatible
with the recent measurements \cite{12)} of the distances to type Ia
supernovae.

In this short note, we discuss the scalar field $\varphi $ as dark matter
and the evolution of matter in the universe after summarizing the Machian
cosmological solution in the (modified) generalized scalar-tensor theory of
gravitation with the varying cosmological constant. Let us start the
following action 
\begin{equation}
S=\int d^{4}x\sqrt{-g}\left\{ -\phi \left[ R+2\lambda (\phi ,\phi _{,\,\mu
}\phi ^{,\,\mu })\right] +16\pi L_{m}-\frac{\omega (\phi )}{\phi }g^{\mu \nu
}\phi _{,\,\mu }\phi _{,\,\nu }\right\} \,,  \label{e1}
\end{equation}
where $\omega (\phi )$ is an arbitrary coupling function and the
cosmological constant is given as 
\begin{equation}
\lambda (\phi ,\phi _{,\,\mu }\phi ^{,\,\mu })\equiv \frac{1}{2}\lambda
_{1}\phi _{,\,\mu }\phi ^{,\,\mu }/\phi ^{2}\,.  \label{e2}
\end{equation}
The action Eq.(\ref{e1}) is equivalent to that of the generalized
scalar-tensor theory without the cosmological constant after replacing $%
\omega (\phi )\rightarrow \left[ \omega (\phi )+\lambda _{1}\right] $.

The energy-momentum tensor $T_{\mu \nu }$ of matter for the perfect fluid is
in general described as 
\begin{equation}
T_{\mu \nu }=-pg_{\mu \nu }-(\rho +p)u_{\mu }u_{\nu }\,,  \label{e3}
\end{equation}
and the conservation law of the energy-momentum $T_{;\nu }^{\mu \nu }=0$
gives the equation of continuity $\dot{\rho}+3\left( \dot{a}/a\right) \left(
\rho +p\right) =0$. Assuming the barotropic equation of state 
\begin{equation}
p(t)=\gamma \rho (t)\,,\;\;-1\leqq \gamma \leqq 1/3\,,  \label{e4}
\end{equation}
we get 
\begin{equation}
\rho (t)a^{n}(t)=const\,,  \label{e5}
\end{equation}
where $n=3(\gamma +1)$.

We are interested in the closed model ($k=+1$) and the line element for the
Friedmann-Robertson-Walker metric is 
\begin{equation}
ds^{2}=-dt^{2}+a^{2}(t)[d\chi ^{2}+\sin ^{2}\chi (d\theta ^{2}+\sin
^{2}\theta d\varphi ^{2})]\,.  \label{e6}
\end{equation}
We obtain the independent field equations 
\begin{eqnarray}
\frac{3}{a^{2}}\left( \dot{a}^{2}+1\right) &=&\frac{\left[ \omega (\phi
)+\lambda _{1}\right] }{2}\left( \frac{\dot{\phi}}{\phi }\right) ^{2}+\frac{%
\ddot{\phi}}{\phi }+  \nonumber \\
&&+\frac{8\pi \rho }{\phi }-\frac{8\pi \left( \rho -3p\right) }{3+2\left[
\omega (\phi )+\lambda _{1}\right] }\frac{1}{\phi }  \label{e7}
\end{eqnarray}
and 
\begin{equation}
\ddot{\phi}+3\frac{\dot{a}}{a}\dot{\phi}=\frac{1}{3+2\left[ \omega (\phi
)+\lambda _{1}\right] }\left[ 8\pi \left( \rho -3p\right) -\frac{d\omega
(\phi )}{d\phi }\dot{\phi}^{2}\right] \,.  \label{e8}
\end{equation}

We require the particular coupling function 
\begin{equation}
\omega (\phi )\equiv \frac{\eta }{\xi -2}\,,  \label{e9}
\end{equation}
where $\xi =1-3\gamma $ or$\ \xi =4-n$, $\eta $ is a constant, and this gives%
$\ d\omega /d\phi =0$. We introduce another scalar function $\Phi (t)$ by 
\begin{equation}
\phi (t)=\frac{8\pi }{3+2\left[ \omega (\phi )+\lambda _{1}\right] }\Phi (t)
\label{e10}
\end{equation}
for the Machian solution satisfying $\phi =O(\rho /\omega )$ and the same
relation between $\dot{\phi}(t)$ and $\dot{\Phi}(t)$, which is guaranteed
for the very slow time-variation or the large value of $\omega $. From Eqs.(%
\ref{e7}) and (\ref{e8}), we find that the expansion parameter has the form 
\begin{equation}
a(t)\equiv A(\omega )\alpha (t)\,,  \label{e11}
\end{equation}
where 
\begin{equation}
\frac{3}{A^{2}(\omega )}=\left| \frac{\omega (\phi )}{2}+B\right| \,.
\label{e12}
\end{equation}

Thus the field equations has a solution 
\begin{equation}
\Phi (t)=\zeta \rho (t)t^{2}\,,\;\;\alpha (t)=bt\,,  \label{e13}
\end{equation}
with 
\begin{equation}
\zeta =1/(\xi -2)\,,  \label{e14}
\end{equation}
\begin{equation}
b=\left\{ 
\begin{array}{l}
(4-\xi ^{2})^{-1/2}\,,\;\;for\;\omega /2+B<0\;and\;0\leqq \xi <2 \\ 
(\xi ^{2}-4)^{-1/2}\,,\;\;for\;\omega /2+B>0\;and\;2<\xi \leqq 4\,,
\end{array}
\right.  \label{e15}
\end{equation}
and 
\begin{equation}
B=\frac{\lambda _{1}}{2}-\frac{3}{(\xi -2)(\xi +2)}\,.  \label{e16}
\end{equation}
There is a discontinuity at $\xi =2$ and we restrict to the range $0\leqq
\xi <2$ owing to the evolutionary continuity from $\xi =0$ (the radiation
era) and $\xi =1$ (the dust-dominated era).

Both the coupling function $\omega (\phi )=\eta /(\xi -2)$ for $0\leqq \xi
<2 $ and the constraint $\omega (\phi )/2+B<0$ lead necessarily to $\eta >0$
and $\omega <0$. We require $\eta >3$ to avoid the singularity ($\omega
<-3/2 $). We find the expansion parameter $a(t)$ and its asymptotic form
when $\xi \rightarrow 2$ 
\begin{equation}
a(t)=\left[ 6/f(\xi )\right] ^{1/2}t\cong \sqrt{3/(2\eta -3)}\,t\,,
\label{e17}
\end{equation}
where the quadratic function $f(\xi )\equiv \lambda _{1}(\xi -2)(\xi
+2)+\eta (\xi +2)-6$. The value $\eta =3$ gives the expansion parameter$\
a(t)=t$ (light velocity) when $\xi \rightarrow 2$. If we require $\lambda
_{1}<0$ and $0\leqq -\eta /\lambda _{1}<2$, the universe (the closed space)
shows the slowly accelerating expansion for the period $-\eta /\lambda
_{1}\leqq \xi <2$.

We find the scalar field $\phi (t)$ and its asymptotic form when $\xi
\rightarrow 2$%
\begin{equation}
\phi (t)=\frac{8\pi \rho (t)t^{2}}{\left( 3+\lambda _{1}\right) (\xi
-2)+2\eta }\cong \frac{4\pi \rho (t)t^{2}}{\eta }\rightarrow const\,,
\label{e18}
\end{equation}
which converges to a definite and finite constant in the limit. As the
parameter $\xi \rightarrow 2$, the coupling function $\omega (\phi )$
diverges to $-\infty $ and the gravitational constant approaches dynamically
to $G_{\infty }$. The cosmological constant $\lambda (t)$ decreases rapidly
in proportion to $t^{-2}$\ as the universe expands and converges to zero
when $\xi \rightarrow 2$: 
\begin{equation}
\lambda (t)=\frac{\lambda _{1}}{2}\left( \frac{\dot{\phi}}{\phi }\right)
^{2}\propto \frac{(\xi -2)^{2}}{t^{2}}\rightarrow 0\,.  \label{e19}
\end{equation}
When $\xi \rightarrow 2$ ($t\rightarrow +\infty $), the abnormal term $%
\omega (\phi )a^{2}\left( \dot{\phi}/\phi \right) ^{2}$ vanishes and the
generalized scalar-tensor theory of gravitation reproduces the correspondent
solution of general relativity with the same energy-momentum tensor, that
is, the Friedmann universe with $\ddot{a}(t)=0$, $\lambda (t)=0$, and $%
p=-\rho /3$.

Let us introduce a scalar field $\varphi $ as dark matter to complete the
physical evolution of matter in the universe, of which the Lagrangian is
given as 
\begin{equation}
L_{\varphi }=-(1/2)g^{\mu \nu }\varphi _{,\,\mu }\varphi _{,\,\nu
}+V(\varphi )\,.  \label{e20}
\end{equation}
The energy-momentum tensor $T_{\mu \nu }^{\varphi }$ derived from this
Lagrangian is 
\begin{equation}
T_{\mu \nu }^{\varphi }=-\varphi _{,\,\mu }\varphi _{,\,\nu }+(1/2)g_{\mu
\nu }\varphi _{,\,\lambda }\varphi ^{,\,\lambda }+g_{\mu \nu }V(\varphi )\,,
\label{e21}
\end{equation}
and the field equation of the scalar field $\varphi (t)$ is 
\begin{equation}
\ddot{\varphi}+3\frac{\dot{a}}{a}\dot{\varphi}+\frac{dV}{d\varphi }=0\,.
\label{e22}
\end{equation}

Let us require a scalar potential $V(\varphi )$ (see \cite{13)}): 
\begin{equation}
V(\varphi (a))\equiv V_{0}/a^{2}>0\,.  \label{e23}
\end{equation}
Taking this potential and the expansion parameter $a(t)=Abt$ into account,
we find a solution of Eq.(\ref{e22}) 
\begin{equation}
\varphi (a)=\left( \sqrt{V_{0}}/Ab\right) \ln a\,,  \label{e24}
\end{equation}
and thus we obtain the exponential potential of the scalar field $\varphi $ 
\begin{equation}
V(\varphi )=V_{0}\exp \left[ -\left( Ab/\sqrt{V_{0}}\right) \varphi \right]
\,.  \label{e25}
\end{equation}

The density of the scalar field $\varphi $ is given as 
\begin{equation}
\rho _{\varphi }=(1/2)\dot{\varphi}^{2}+V(\varphi )=(3/2)\left(
V_{0}/a^{2}\right) \propto t^{-2}\,,  \label{e26}
\end{equation}
and the pressure is 
\begin{equation}
p_{\varphi }=(1/2)\dot{\varphi}^{2}-V(\varphi )=-(1/2)\left(
V_{0}/a^{2}\right) \,.  \label{e27}
\end{equation}
Thus, we get the relation $p_{\varphi }=-(1/3)\rho _{\varphi }$ and may
regard the scalar field $\varphi $ as a perfect fluid with the barotropic
equation of state Eq.(\ref{e4}). It should be noted that the above Machian
cosmological solution produces no changes by introducing the scalar field $%
\varphi $. Let us suppose that the total matter of the universe consists of
the radiation (density $\rho _{r}$, $\gamma =1/3$), the dust-matter (density 
$\rho _{m}$, $\gamma =0$), and the scalar field $\varphi $ (density $\rho
_{\varphi }$, $\gamma =-1/3$) for simplicity:\ 
\begin{equation}
\rho =\rho _{r}+\rho _{m}+\rho _{\varphi }\,.  \label{e28}
\end{equation}
The densities $\rho _{r}$, $\rho _{m}$, and $\rho _{\varphi }$ satisfy the
conservation laws $\rho _{r}a^{4}=\rho _{r0}a_{0}^{4}$, $\rho _{m}a^{3}=\rho
_{m0}a_{0}^{3}$ , and $\rho _{\varphi }a^{2}=\rho _{\varphi 0}a_{0}^{2}$
respectively, where a subscript $0$ denotes the present value. The total
pressure is 
\begin{equation}
p=p_{r}+p_{m}+p_{\varphi }\,,  \label{e29}
\end{equation}
and satisfies the equation state $p=\gamma (t)\rho $.

We find a time when the total pressure vanishes ($\xi =1$) 
\begin{equation}
a=\sqrt{\rho _{r0}/\rho _{\varphi 0}}\,a_{0}\,.  \label{e30}
\end{equation}
When $\rho _{r}\ll \rho _{m}\ll \rho _{\varphi }$, which is valid enough at
present, we get 
\begin{equation}
\epsilon \equiv 2-\xi \cong \frac{\rho _{m0}}{\rho _{\varphi 0}}\frac{a_{0}}{%
a}\,.  \label{e31}
\end{equation}
Taking $\left| \omega \right| \sim 10^{3}$ \cite{14)} and $\eta \approx 3$
into account, we obtain $\epsilon \sim 10^{-3}$ from Eq.(\ref{e9}), and thus
find $\rho _{m0}/\rho _{\varphi 0}\sim 10^{-3}$ at present. If we adopt $%
\rho _{r0}/\rho _{m0}\sim 10^{-3}$, we get $a\sim 10^{-3}a_{0}$ for the time 
$\xi =1$. It is interesting that it is the time when the
hydrogen-recombination finished in the universe. As the critical density $%
\rho _{c}\sim 10^{-29}\,g.cm^{-3}$ leads to the valid gravitational constant 
\cite{10)}, we may adopt $\rho _{\varphi 0}\sim \rho _{c}$. In this case, we
find $\rho _{m0}/\rho _{\varphi 0}\sim 10^{-2}$ and get $\epsilon \sim
10^{-2}$, which gives $\left| \omega \right| \sim 10^{3}$ with $\eta \approx
10$. When the expansion parameter $a(t)\rightarrow +\infty $ ($t\rightarrow
+\infty $), the parameter $\xi $\ surely approaches to $2$, the final state
of the universe.

Our universe started (classically) from the Big Bang with $\xi =0$ (the
radiation era), passed the dust-dominated era ($\xi =1$) rapidly in the
early stage, and has been staying the negative pressure era ($\xi \approx 2$%
) for the almost all period of $10^{10}yr$. The barotropic state of the
universe has been varying extremely slowly from $\xi =0$ to $\xi \approx 2$.
At $t\sim 5\times 10^{9}\,yr$, the parameter $\epsilon $ was $2\times
10^{-3} $, and thus the gravitational constant was almost $G_{\infty }$ as
the same value at present. This situation saves the evolution of life in the
Earth in relation to the nuclear fusion of the Sun. Finally, the universe
will approach to the state of $\xi =2$ ($\gamma =-1/3$) as it expands for
ever.

As a result of the barotropic evolution of matter in the universe, the
coupling function $\omega (\phi )$ diverges to the minus infinity when $%
t\rightarrow +\infty $ and the gravitational constant approaches dynamically
to the constant $G_{\infty }$. The cosmological constant $\lambda (t)$
decreases in proportion to $t^{-2}$\ and converges to zero when $%
t\rightarrow +\infty $. The generalized scalar-tensor theory of gravitation
reduces dynamically to general relativity. The Machian cosmological model
approaches to the Friedmann universe in general relativity with $\ddot{a}%
(t)=0$, $\lambda (t)=0$, and $p=-\rho /3$.

The scalar field $\varphi $ play the most important role in the physical
evolution of the universe and in the problem of dark matter. Our conjecture 
\cite{11)} on the barotropic history of the universe was proved. However,
the next puzzle arises: Why does the scalar field $\varphi $ have the
exponential potential like Eq.(\ref{e25})? We also conceive another
conjecture: The coupling function $\omega $ is rather derived from the
scalar field $\varphi $.\newline
\newline
\textbf{Acknowledgment}

The author is grateful to Professor Carl Brans for helpful discussions and
his hospitality at Loyola University (New Orleans) where this work was done.
He would also like to thank the Nagasaki Prefectural Government for
financial support.


\begin{thebibliography}{99}
\bibitem{1)}  C.Brans and R.H.Dicke, Phys. Rev. 124, 925 (1961).

\bibitem{2)}  N.Banerjee and S.Sen, Phys. Rev. D56, 1334 (1997).

\bibitem{3)}  V. Faraoni, Phys. Rev. D59, 084021 (1999).

\bibitem{4)}  A.Miyazaki, gr-qc/0012104, 2000.

\bibitem{5)}  A.Miyazaki, gr-qc/0101112, 2001.

\bibitem{6)}  A.Miyazaki, gr-qc/0102003, 2001.

\bibitem{7)}  P.G.Bergmann, Int. J. Theor. Phys. 1, 25 (1968).

\bibitem{8)}  R.V.Wagoner, Phys. Rev. D1, 3209 (1970).

\bibitem{9)}  K.Nordtvedt, Astrophys. J. 161, 1059 (1970).

\bibitem{10)}  A.Miyazaki, gr-qc/0102105, 2001.

\bibitem{11)}  A.Miyazaki, gr-qc/0103003, 2001.

\bibitem{12)}  S.Perlmutter, M.S.Turner, and M.White, Phys. Rev. Lett. 83,
670 (1999).

\bibitem{13)}  T.Matos, F.S.Guzm\'{a}n, and L.A.Ure\~{n}a-L\'{o}pez,
astro-ph/9908152.

\bibitem{14)}  X.Chen, M.Kamionkowski, Phys. Rev. D60, 104036 (1999).

\bibitem{15)}  O.Bertolami and P.J.Martins, Phys. Rev. D61, 064007 (2000).
\end{thebibliography}
\end{document}